\documentclass[12pt,preprint]{aastex}
\shorttitle{Planets in 47 Tucanae}
\shortauthors{Weldrake et al.}
\begin{document}
\title{An Absence of Hot Jupiter Planets in 47 Tucanae: Results of a Wide-Field Transit Search.}

\author{David T F Weldrake} 
\affil{Research School of Astronomy and Astrophysics, Australian National University, Mount Stromlo Observatory, Cotter Road, Weston Creek, ACT 2611, Australia}
\email{dtfw@mso.anu.edu.au}

\author{Penny D Sackett}
\affil{Research School of Astronomy and Astrophysics, Australian National University, Mount Stromlo Observatory, Cotter Road, Weston Creek, ACT 2611, Australia}
\email{psackett@mso.anu.edu.au}

\author{Terry J Bridges}
\affil{Anglo-Australian Observatory, P.O. Box 296, Epping. NSW, 1710, Australia}
\affil{Current Address: Department of Physics, Queen's University, Kingston, Ontario. 
\linebreak K7L 3N6, Canada}
\email{tjb@astro.queensu.ca}
\and

\author{Kenneth C Freeman} 
\affil{Research School of Astronomy and Astrophysics, Australian National University, Mount Stromlo Observatory, Cotter Road, Weston Creek, ACT 2611, Australia}
\email{kcf@mso.anu.edu.au}

\begin{abstract}
This paper presents the results of a comprehensive wide field search for transiting ``Hot Jupiter'' planets (gas giant planets with an orbital period 1d$\le$P$\le$16d) in the globular cluster 47 Tucanae. Motivated by the detection of the transit in HD209458 and the apparent lack of planetary detections in the core of 47 Tuc by Gilliland and coworkers, this work further addresses the question of giant planet frequency in 47 Tuc by observing, from the ground, a 52$'$$\times$52$'$ field centered on the cluster. Hence this work is most sensitive to the uncrowded outer regions, where the stellar densities are significantly lower than the core, and concentrates on 21,920 main sequence stars within 2.5 magnitudes of the cluster turnoff (and hence approaching solar in mass) Our work comprises the largest ground-based transit search of a globular cluster to date, incorporating a 33-night time-series which allows us excellent sensitivity to detect Hot Jupiter planets.

Detailed Monte Carlo simulations incorporating the actual temporal sampling and photometric precision of the data predict that seven planets with an orbital period range of 1$-$16d should be present in our dataset, if 47 Tuc has the same planetary frequency observed in the solar neighbourhood. A detailed search utilizing a matched filter algorithm, developed specifically for this project found no transit events. This 3.3$\sigma$ result is consistent with the HST cluster core null detection of Gilliland and coworkers. Our result indicates that system metallicity rather than crowding is the dominant effect inhibiting Hot Jupiter formation in this environment. The 33-night dataset used for this result also led to the detection of 100 variable stars, including 69 new discoveries, which have been presented in a companion paper.

\end{abstract}

\section{Introduction}
Using transit photometry to search for close-in extrasolar giant planets (Hot Jupiters) has recently yielded success. A planet of Jupiter's radius orbiting at $<$0.1AU will cause a 1-2$\%$ dimming of $\sim$2 hours duration in a solar-type main sequence parent star, if its orbital inclination is sufficiently edge-on as viewed from Earth. Transits allow the determination of a number of important system characteristics, including the orbital period, an estimate for the orbital inclination and the planetary radius. When coupled with radial velocity observations, an accurate measure of the planetary mass is determined. Hence, such transit-determined parameters are important for theoretical studies of extrasolar planet formation and evolution. 

The availability of moderate aperture telescopes over long contiguous observing windows has allowed many transit candidates to be discovered in massive campaigns. HD209458b was the first planet to be observed with the transit method \citep{Char2000,Henry2000}, being first discovered using radial velocity techniques \citep{Mazeh2000}. This Hot Jupiter was later followed up by \citet{Brown2001} using HST/STIS, who achieved a photometric precision of 0.1 mmag and a temporal resolution of 80s. Using this dataset, the planetary radius was estimated to be R$_{p}$=1.35$\pm$0.06R$_{Jup}$. With knowledge of the planetary mass (M$_{p}$=0.69$\pm$0.05M$_{Jup}$), an estimate was made for the density (0.35 g cm$^{-3}$). This planet seems to have an unusually large radius for it's mass. \citet{Bar2003} discusses the need for an additional source of heat to account for these radius measurements, whereas \citet{Burrows2000} suggests that the cause is due to high residual entropy caused by the planet's formation very close to the host star. Recent transit discoveries have suggested that the extended atmosphere of HD209458b is somewhat unusual \citep{Burrows2004}, and is likely caused by the increased stellar flux retarding the contraction of an extended planetary radius from the time of formation.

Using the 1.3m Warsaw telescope in Chile, the OGLEIII group originally discovered 46 transit candidate systems towards the Galactic Bulge \citep{Udal2002}. The transiting signals in these systems could be caused by planets, brown dwarfs, stellar companions or blended eclipsing binaries. A total of 137 candidates have been presented more recently by \citet{Udal2003}, and are the subject of vigorous follow-up programs to determine their likely nature. From these candidates, OGLE-TR-56b became the first planet discovered using the transit method \citep{Kon2003a} and confirmed with radial velocity data that measured the planetary mass \citep{Torres03,Kon2003b}. The estimates for the planetary parameters of OGLE-TR-56b are M$_{p}$=1.45$\pm$0.23M$_{Jup}$ and R$_{p}$=1.23$\pm$0.16R$_{Jup}$, which produce a density of 1.0 g cm$^{-3}$, significantly higher than that of HD209458b. Very recently, two new transiting planets have been confirmed by \citet{Bouchy04}, OGLE-TR-113b and OGLE-TR-132b. Of these, OGLE-TR-113b was independently verified by \citet{Kon2004}. Both of these new discoveries have very short orbital periods ($\sim$1.2d), much shorter than the apparent 3d cutoff seen in the distribution of radial velocity Hot Jupiter discoveries \footnote{Using data taken from the Extrasolar Planet Encyclopedia http://www.obspm.fr/encycl/encycl.html}, either indicating the presence of a new class of close-in `Very Hot Jupiters', or revealing a selection effect. Either way, these objects present interesting targets for theoretical work. Very recently, \citet{Pont2004} announced the planetary nature of OGLE-TR-111, and \citet{Alonso2004} announced TrES-1, the first planet transit detected via the TrES multi-site campaign.

Many groups are searching for transiting planets. For example \citet{Moch2003} have undertaken the Planets in Stellar Clusters Extensive Search (PISCES) targeting selected open clusters, and \citet{Hidas2003} have undertaken a wide field search for transits in the galactic plane, and expect about seven Hot Jupiter discoveries a year. Further information on the status and prospects of current selected transit searches can be found in the review by \citet{Horne2003}.

Globular clusters are an excellent laboratory in which to search for transiting planets. 47 Tuc is one of the brightest globulars in the sky, and is well-placed for observation from the Southern Hemisphere. Additionally, it is among the closest globulars (around 4.7Kpc \citep{Harris96}), and thus well resolved by small telescopes, and contains bright main sequence objects for study (V=17.0+). With a suitably wide field instrument it is possible to observe the whole of the cluster in a single exposure, capturing tens of thousands of target stars in a single field of view. This maximises the temporal coverage, increasing the probability of a successful detection, and raises the statistical significance of any null result. The cluster has excellent observability at its respective season from the south, and is located far from the path of the Moon. 

Motivated by the detection of the transit in HD209458 and by the apparent lack of planetary detections in the inner, crowded region of the globular cluster 47 Tucanae by \citet{Gil2000}, we further address the question of which conditions are necessary for giant planet frequency by observing, from the ground, a very wide field centered on 47 Tuc. We concentrate on transit detection involving main sequence stars close to the cluster turnoff (MSTO), and hence on parent stars approaching solar mass. This work complements that of \citet{Gil2000}, who used WFPC2 on HST to search for transits on an ensemble of $\sim$34,000 main sequence stars in a field centered on the core of 47 Tuc, by effectively performing a ``control experiment'' outside the core. We search for the same transit signatures in the uncrowded outer regions of the cluster, utilising a field of view $\sim$250 times larger, to examine whether the paucity of Hot Jupiters in the core is due solely to the density of the stellar environment. Ours is the largest ground-based transit search of a globular cluster to date. We compare our results to the frequency of known Hot Jupiters in the solar neighbourhood.

The frequency of planetary systems appears to be dependent on the metallicity of the host star \citep{Gon1997,L2000,Fischer2003}, with stars of low metallicity having a lower probability of planetary formation. Globular clusters allow a direct test to be made on this dependence. These clusters are among the oldest of astronomical objects, and their stars are typically therefore of very low metallicity. Our target, while metal-rich as far as globulars are concerned (47 Tuc [Fe/H]=-0.76 \citep{Harris96}), is still metal poor compared to the Solar Neighbourhood. Furthermore, globular clusters allow the effect of stellar crowding on planetary formation to be studied. Such crowding may disrupt protoplanetary disks and harden typical planetary systems. Stellar encounters would be more commonplace in the cluster core, sampled by the \citet{Gil2000} HST survey, but much less common in the uncrowded outer regions where this experiment is most sensitive. By considering the timescale needed for planet formation and migration, which is generally regarded to be short ($<$10$^{5}$yrs)\citep{Boss2001}, along with the relaxation time, which increases with increasing radius from the core, the region chosen for a transit search could have a large bearing on the probability of a detection. Short crossing times would have a larger effect on circumstellar material at early times and could inhibit planet formation. We test this by sampling the whole of the visible cluster face (except the inner 6$'$) out to $\sim$60$\%$ of the cluster tidal radius of 45$'$.9 \citep{Leon2000} in one exposure. 

Section 2 describes the observations and data reduction, along with notes on the method used to produce the photometry and astrometry. Section 3 details the expectations for our transit search if the planet frequency in 47 Tuc is similar to the solar neighbourhood, with a brief description of the algorithm developed to carry out the search, which is presented in full in a companion paper. Section 4 describes the results of our search and its implications. Section 5 contains the summary and conclusions.

\section{Observations and Data Reduction}
Full details of the observations, data reduction, photometry and astrometry have been given in \citet{Weld2004a}, hereafter WSBFa, a companion paper presenting the variable star catalogue produced from the same dataset, and so these aspects are only summarised here. Using the Australian National University (ANU) 40-inch (1m) telescope at Siding Spring Observatory, 47 Tucanae was observed for a total of 33 nights from 2002 August 22 to 2002 September 24. Approximately 80$\%$ of the time was useful for the transit search, with mean seeing of 2.2 arcsec. The Wide Field Imager (WFI) was used, consisting of a 4$\times$2 array of 2048$\times$4096 pixel back-illuminated CCDs, and providing a field of view 52$'$ on a side, with a 0.38$''$ pixel detector scale. This field of view allows the whole of the visible extent of the cluster (out to 60$\%$ of the cluster tidal radius) to be imaged in a single exposure. The field layout can be seen in Fig.1 of WSBFa.

In order to achieve a significant increase in S/N for any given exposure time while keeping atmospheric dispersion to an undetectable level, a single broadband filter covering Cousins V and R was used. Temporal resolution was maximised as much as possible  with 300s exposures taken on average every six minutes for around 10 hours per night. The dataset in total comprises 1220 images of the same field centered on 47 Tuc, which have been used to produce time-series lightcurves. Image offsets were minimised at the telescope, so that the database of stars is maximised and the effects of CCD gaps reduced. Any images containing satellite trails or taken in bad conditions were discarded at the telescope. In total our database contains 109,866 stellar lightcurves with a 33-night temporal extent..
 
Image reduction was undertaken using the MSCRED package within IRAF\footnote{IRAF is distributed by National Optical Observatories, which is operated by teh Association of Universities for Research in Astronomy, Inc., under cooperative agreement with the National Science Foundation.}. Standard reduction practices were employed, including region trimming and overscan correction, bias correction, flat-fielding and dark current subtraction.

\subsection{Photometry}
We produced the time-series lightcurves using a Difference Image Analysis (DIA) method originally described as an optimal PSF matching algorithm by \citet{AL98}, and later modified by \citet{Woz2000} for use in detecting microlensing events. Rather than producing the total flux per image, this method measures the difference in flux of stellar objects between individual dataset frames, and a template frame. Lightcurves are generated in a linear flux unit, from which the constant flux of that object on the template frame has been subtracted using a specified PSF. This is made to be equal across the whole dataset, reducing systematic effects. We estimated the template flux of each star that was visible in our best seeing image (FWHM$=$0.8$''$), and then converted the linear flux unit to differential magnitude units for the main analysis.

The details of our photometric procedure and uncertainties are given in WSBFa. Of importance here is that the field contains a large change in crowding between the four inner and four outer CCDs, with a corresponding change in photometric uncertainty at a given magnitude limit. We have sufficient photometric quality to detect a 1.5-2.5$\%$ dip, typical of transiting Hot Jupiter planets, down to V=18.5 and V=19.0 in crowded inner and uncrowded outer regions of 47 Tuc, respectively. Our rms uncertainty is 0.04 mag for V$=$19.5 stars, allowing us to probe at least the brightest two magnitudes of the cluster main sequence for any orbiting giant planets. The ability of our algorithm to detect transits even in the presence of substantial noise has a direct bearing on the faint magnitude limit of our search, as discussed in the following section.

\section{Transit Expectations}
For this work, we have defined a Hot Jupiter as having an orbital period less than 16d, similar to that of 14d used by \citet{Basri04}, and roughly corresponding to an apparent gap in the orbital period distribution of exoplanets\footnote{http://www.exoplanets.org}. This definition is somewhat arbitrary; our definition comes from the limit of orbital period that can be theoretically detected in our dataset.
  
In order to search for specific features in the photometric time-series, maximum sensitivity can be obtained if the parameters of the features can be anticipated, and a matched filter algorithm used. Here we will describe the calculation of the expected depths and durations of transits in our dataset. Ignoring the effects of stellar limb darkening, the transit depth can be determined from a simple ratio of the planetary and stellar radii, as derived by considering the maximum fractional change in the observed flux:
\begin{center}
\begin{equation}
\mathrm{Depth} \sim \left( \frac{R_{p}}{R_{\ast}} \right) ^2
\end{equation}
\end{center}
\noindent where $R_{p}$ is the radius of the planet, and $R_{\ast}$ is the radius of the stellar host. Estimates can be made for the mass and radius of a star in 47 Tuc at any given V and V-I from \citet{VB2001}, and are tabulated in Table1 for stars of 17.0$\le$V$\le$19.5 (corresponding $M_{V}$ from 3.6 to 6.1), assuming a 47 Tuc distance modulus of 13.4. 

As 47 Tuc is a very old object and with an [Fe/H] of -0.7, the stellar mass at the Main Sequence Turn Off (MSTO) is estimated to be $\sim$0.9M$_{\odot}$ \citep{VB2001}, corresponding to luminosities $\sim$2.5L$_{\odot}$ and radius 1.5$_{\odot}$. It therefore follows that the stars sampled for transits in this paper will have a maximum mass of this limit (0.9M$_{\odot}$). Stars beyond the cluster turnoff (and not on the red giant branch) are physically larger and would therefore exhibit lower amplitude transits. Stars on the main sequence of 47 Tuc have smaller radii than solar-mass stars in the solar neighbourhood, leading to a deeper transit signature, which makes them easier to detect.

The MSTO for 47 Tuc is at V$\sim$17.2, taken from the work of \citet{Hesser87}, and consistent with that found by WSBFa from their Colour-Magnitude diagram, reproduced in Fig.\ref{47Tuc_tr_depths}. We take this as the bright stellar limit in our search of 47 Tuc stars for Hot Jupiter companions. The lower limit depends on the ability of our detection algorithm to detect the deepest transits that could conceivably be caused by a planetary companion. From Monte Carlo simulations involving modelled transits of varying depth, we found that our algorithm can detect 3$\%$ depth (0.03 mag) transits in V$=$19.5 stars, which typically have an rms photometric uncertainty of 0.04 mag, forming the lower photometric limit to our search. A total of 21,920 stars lie between these bounds; these are the stars we search for Hot Jupiters.

In order to determine the expected duration of a planet transit as a function of V magnitude along the main sequence of 47 Tuc, we must consider the relationship between the length of transit ($\tau_{tran}$) and stellar parameters (see, for example \citet{Gil2000}:
\begin{center}
\begin{equation}
\tau_{tran} = 1.412M_{\ast}^{-1/3}R_{\ast}P_{orb}^{1/3}
\end{equation}
\end{center}
\noindent where $M_{\ast}$ and $R_{\ast}$ are the stellar mass and radius in solar units where $P_{orb}$ is the orbital period of the planet in days, and $\tau_{tran}$ is also given in days. This relationship includes a $\pi$/4 reduction in transit duration, and hence is the expected duration for a planet which crosses the `average' chord length of the stellar disk. A centrally-crossing transit has a duration 1.273 times longer and a grazing transit duration is 0.66 times shorter. Using Eq.1 and 2 and the results given in Table 1, we now calculate the transits depths and durations (assuming the `average' chord length) expected for any Hot Jupiters in our dataset. These are shown schematically in Fig.\ref{47Tuc_tr_depths}, where we indicate the transit depth of a 1.3 $R_{J}$ planet, with an assumed orbital period of 3.3d, (the most typical orbit for such a planet in the Solar Neighbourhood\footnote{http://www.obspm.fr/encycl/encycl.html}) as a function of V magnitude. The upper and lower limits of the transit search range are defined by the location of the MSTO and the point where our photometric uncertainty is poorer than 0.04 mag (rms).

Hot Jupiter transit depths are expected to vary from 0.007 mag to 0.031 mag over our search range, with the shallower transits having a longer duration than the deeper, due to the larger radii of the parent stars. These two factors have a role to play in determining the visibility of a transit in the dataset. It is clear that a transit can only occur if the orbital plane is inclined at a suitably critical viewing angle. This inclination $\it{i}$ must therefore satisfy:
\begin{center}
\begin{equation}
\alpha \cos i = R_{\ast}+R_{p} 
\end{equation}
\end{center}
\noindent where $\alpha$ is the planetary semi-major axis, $\it{i}$ is the orbital inclination, $R_{\ast}$ is the stellar radius, and $R_{p}$ is the radius of the orbiting planet. However, $\cos$ $\it{i}$ can take on any random number from 0 to 1, so to derive a transit probability, P$_{tran}$ for a large sample of randomly oriented targets, this must be taken into account by integrating over all possible orientations. We take P$_{tran}$ to be given by \citet{Sackett1999}:
\begin{center}
\begin{equation}
P_{tran} = \frac{\int_{0}^{(R_{\ast}-R_{p})/a}d(\cos i)}{\int_{0}^{1}d(\cos i)} = \frac{R_{\ast}-R_{p}}{\alpha} \sim 0.87 \big(\frac{R_{\ast}}{\alpha}\big)
\end{equation}
\end{center}
\noindent We have taken the upper limit to be appropriate for a grazing transit, that is, we have assumed that the limit of detection is defined when the outer edge of the planetary disk just grazes the edge of the stellar disk. The factor of 0.87 is derived using an assumed planetary radius of 1.3$R_{J}$. Using this relationship, the transit probability of a planet of 1.3R$_{J}$ orbiting a solar radius primary, typical of the stellar radius at the centre of our search range is plotted as a function of orbital period ($P=1-16d$) and is presented in Fig.\ref{transprobs}. The transit probability drops dramatically as period increases: at 2d orbital period the probability is 13.0$\%$, whereas at 8d, the probability drops to 5.2$\%$. An orbital period of 16d has been chosen as the upper limit of the search as any planet at this period is at the theoretical limit of displaying three detectable (and hence confirmed) transits within our 33-night observing window. It is of course possible to have only one or two transits visible in the data. However, in our search these are wholly caused by systematic effects as they all occurred at the same times in the dataset and were much more common on stars in the most crowded regions of the data with common periodicities.

\subsection{Method for transit detection.}
 We employed a Matched Filter Algorithm to search for the periodic transit signature in our time series lightcurves. Full details of the method used are given in our second companion paper, \citet{Weld2004b}, hereafter WS. The matched filter method was first suggested for use in transit detection by \citet{Jenk96}, has been described as the best method for transit searches in the literature \citep{Ting03a,Ting03b}, and has been used as the method of choice for several transit searches in recent years \citep{Gil2000,Bruntt03}. Our MFA was designed and produced specifically for this project and has been extensively tested with Monte Carlo simulations to test for and maximize the recoverability of modelled transits while keeping the false detection rate to an acceptable level. These tests involve the addition of false transits of various depths, durations and periods onto the actual time-series dataset. Thus, the actual time sampling and photometric uncertainty of the data is taken into account by our Monte Carlo recoverability simulations.

The method as implemented assumes a simple square-well model, which is a valid assumption when searching for a signal very close to the noise level. It searches for multiple transits spanning a large range of periods and transit start times. It was found from Monte Carlo simulations of our transit detection algorithm that the data must be analysed in two regions of the CMD, as they both have slightly different detection criteria. These criteria are described fully in \citet{Weld2004b}, which deals with describing the MFA in detail. The resultant two bins are defined on Fig.\ref{47Tuc_tr_depths}. Depths and durations of model transits are these expected for Hot Jupiters accounting for different positions the transit could have on the stellar disk as shown in Fig.\ref{47Tuc_tr_depths}. Detailed testing revealed slightly different model depths within this range did not change the transit recoverability level to any significant degree, as the degree of scatter in the data is comparable to the transit depth.

A total of 1,133,255 different transit models were compared to each sampled lightcurve and a cross-correlation function (CCF) value created for each. A detection is indicated by a strong spike in the histogram of all CCF values. The results of the transit recoverability tests show that the code is very efficient at detecting transit signatures at all sampled orbital periods. For stars with photometric uncertainties $\le$0.02 mag (rms), the recoverability spans 95$\%$ to 35$\%$ for orbital periods of 1-16d; including the recoverability of stars with rms uncertainty of 0.02-0.04 mag yields a weighted mean recoverability of 85$\%$ to 25$\%$ for the complete sample of stars over the same range of period.

\subsection{Expected Number of Hot Jupiter Detections}
Using the recoverability function $\it{R}$ of the Monte Carlo simulations, seen in Fig.11 of \citet{Weld2004b}, we estimate the number of Hot Jupiter planets that should be detectable in our dataset, allowing a statistical significance to be placed on any null result. In order to do this, a number of factors need to be taken into consideration, including intrinsic frequency of Hot Jupiters, the total number of stars analysed, the geometric transit probability at any particular given orbital period and the recoverability at that period delivered by the Matched Filter Algorithm. 

We begin by assuming that the frequency of Hot Jupiters, $\it{O}_{HJ}$, in 47 Tuc is the same as in the solar neighbourhood, which has been estimated to be 0.8$\%$ \citep{Butler2000,Basri04}. This is then multiplied by the total number of main sequence stars available for analysis, $\it{n}_{\ast}$, (21,920, 8,080 with rms $\le$0.02 mag, and 13,840 with 0.02$<$rms$<$0.04 mag) to give the total number of Hot Jupiters that should be present in our dataset. The total number of Hot Jupiters are not randomly distributed in orbital period. Using the period distribution\footnote{http://www.obspm.fr/encycl/encycl.html} of Hot Jupiters to date we split them into period bins of 0.5d width to separately calculate the expected number of detections in each period bin $\it{i}$ and then sum these to arrive at:
\begin{center}
\begin{equation}
N_{HJ} = \sum_{i=1}^n N_{HJ} (P_{i}) = (n_{\ast}\times O_{HJ})\times \bigg\{ (f_{HJ})_i \times(P_{tran})_i \times(R_i) \bigg\}
\end{equation}
\end{center}
\noindent where $N_{HJ}$ is the total number of detectable Hot Jupiters (HJ) present in the dataset, $(n_{\ast}\times O_{HJ})$ is the total number of Hot Jupiters assumed to be present in the dataset, $f_{HJ}$$_{i}$ is an estimate of the fraction of total Hot Jupiters expected to have periods in bin $\it{i}$ as estimated from the solar neighbourhood, ($P_{tran}$$_{i}$) is the geometric transit probability for bin $\it{i}$ (Eq.4) and $(R_i)$ is the recoverability delivered by our detection algorithm as determined in the Monte Carlo simulations, as seen in Fig.11 of \citet{Weld2004b}. 
 
The resulting expected distributions are presented in Table 2 and Fig.\ref{expnumbers}. In these calculations, the recent OGLE transit detections \citep{Kon2003a,Bouchy04,Kon2004} have been omitted, as the follow-up classification studies are on going, so the statistics cannot be calculated. Moreover, the OGLE detections were in environments other than the Solar Neighbourhood, and thus may have a different intrinsic Hot Jupiter frequency. Fig.\ref{expnumbers} shows that there should be a total of 7 detectable Hot Jupiters in our dataset, assuming that the formation probability is the same in 47 Tuc as in the solar neighbourhood. Using Poission statistics, a probability of no detections when 7 are expected would be 1/1,097, equating to a 3.3$\sigma$ null significance. 

It is noted that from recent transit discoveries this assumed radius may be anomalously large, and therefore was altered to a radius of 1.0 $R_{J}$. The resultant transit recoverability and expected numbers were not changed to any significant degree.

Planet formation or survivability appears to depend on the metallicity of the host star \citep{Gon1997,L2000,Santos2001,Fischer2003}, with the vast majority of planet host stars being of supersolar metallicity. However, only a few percent of the stars sampled for Hot Jupiter planets have metallicities similar to 47 Tuc \citep{Fischer2003}. Assuming that a better proxy for the frequency of Hot Jupiters in 47 Tuc is given by frequency of such planets around solar neighbourhood stars of similar metallicity, we show with the dotted line in Fig.\ref{expnumbers} that the expected number of planets in the dataset would have to be revised downward to 0.9. Hence, a null result would be consistent with the expected number if the system metallicity is the dominant factor inhibiting planet formation in this environment.

\section{Results}
We ran our transit detection algorithm on the entire dataset of main sequence stars, with photometric rms $\le$0.04 mag using slightly different detection criteria for the two main analysis bins outlined in Fig.\ref{47Tuc_tr_depths}. For the 8,080 stars with rms $\le$0.02 mag, a total of 64 possible ``detections'' were found. These detections were then phase-wrapped to all the significant periods to determine the nature of the periodicity that was seen. Phase wrapping indicated that the vast majority showed no clear pattern, but were instead caused by localised increases in scatter. These small-scale increases in random errors are very difficult to account for in the code when randomly spread in the timeseries, but are easily identified by eye when the data are phase wrapped to the periods singled out by the cross correlation function.

In order for a candidate to be counted as a real transit detection, the lightcurve must pass several stringent criteria. The star must be within the V and V-I limits of the search. It must not be located close to a bright saturated object, as this can cause spurious transit signals in the data. It must also not be located very close to a bad column or other CCD blemishes, as image offsets during observing could cause the star to be periodically placed over these features. The star must have no heavily blended companions that degrade the ability of the photometric code to fit a reasonable point spread function (PSF) to the star. This can be tested easily by checking the $\chi^{2}$ goodness of fit for the PSF. A significantly high $\chi^{2}$ value indicates that the PSF is not being modelled appropriately, resulting in systematic deficiencies in lightcurve photometry. This could be caused by differing colours between the two heavily blended stars, or by other systematic effects introduced by ground-based photometry that are difficult for the photometry code to handle. This was only a problem on very few of those stars with heavily blended neighbours. The candidates were comprehensively searched and those which displayed apparent transit features checked against these criteria. They were also checked with multiples of the detected periods, to see if any of the transits repeated. No signals which could be interpreted as being caused by a planet transit were observed.

Our Matched Filter Algorithm did, however, detect some interesting systematic effects, including a star that was contaminated by a nearby detached eclipsing binary and gave the highest significance detection (9.5$\sigma$) in this sample. The responsible binary was already identified as V39 (P=4.6015d) and published in WSBFa. The contaminated star is shown in Fig.\ref{contamEcB}, and is a good example of the ability of the code to detect low amplitude signals. The main eclipse occurs at Phase ($\Phi$)=0 and $\Phi$=1, with the depth and duration marked. The secondary eclipse can be clearly seen at $\Phi$=0.5. No additional variable stars were identified in our transit search over the variables already identified in WSBFa using another algorithm. As only two binary stars were recovered, it seems that the vast majority of the variables available in this dataset (100 in total, 69 new discoveries) have already been identified and published in our companion paper WSBFa.

Twelve transit-like features were identified in our search that had a common period of 1.66d; one of these is plotted in Fig.\ref{sysdet}. This feature (at $\Phi$=1) has the correct depth for a Hot Jupiter companion, and a duration (1.39 hours) that would be typical of a planet which only grazes the stellar disk. If this detection had not occured at a common periodicity, and at common places in the time-series, this would be a candidate for a grazing Hot Jupiter or eclipsing binary. We present this false detection here as an excellent example of the ability of the detection algorithm to identify even short duration transit-like events, which is typical of worst-case scenario transits where the planetary disk just fits inside the stellar disk.

For the second sample of stars, with larger rms photometric uncertainties of 0.02-0.04 mag, a total of 48 ``detections'' were found in the dataset. Again, these candidates were phase-wrapped at their respective determined periods and the nature of the periodicity seen. After checking in detail for the systematic effects listed above, no transits were observed.

\subsection{Interpretation}
Our null result has high statistical significance (3.3$\sigma$), and is in agreement with the findings of \citet{Gil2000} that the environments of globular clusters (at least 47 Tuc) lead to a dramatic reduction in Hot Jupiter formation probabilities. There are several possible scenarios that can explain this result. 

The main conclusion from the work of \citet{Gil2000} in the core of 47 Tuc was that either metallicity or crowding is the most important factor for significantly reducing the formation of Hot Jupiter planets in this environment. As our search is most sensitive to the uncrowded outer regions of the cluster, where the cluster relaxation time is longer, we would expect crowding to have a much smaller effect on our sample than samples taken from the cluster core. In any event, the work of \citet{Sig1992} shows that the survivability of a planet at 1AU semi-major axis is suprisingly long ($\sim$10$^{8}$yrs) even for conditions typical of the core of 47 Tuc, and for P$\le$5d orbital period planets, the stellar crowding would disrupt and eject only a small fraction from their parent systems. 

\citet{Bonnell2001} undertook a further investigation into the effect of stellar crowding on planet formation. The conclusion was that the disruption of Hot Jupiter planets is unlikely to occur owing to encounters in a stellar cluster, due to the long timescales involved. Planets with long orbital period can however be ejected from their parent system. In the core of such a cluster with densities $>$10$^{4}$ star pc$^{-3}$, even those planetary systems with semi-major axes of 0.1 au are likely to be disrupted and only those systems with semi-major axes $<$0.01 au are likely to survive. Lower densities typical of the halo of the cluster (where this work is most sensitive) will leave wider systems ($<$10 au semi-major axis) intact. 

Additionally, two of the variable stars (V78 and V93) identified in the field of 47 Tuc appear to be orbited by M-dwarfs with orbital periods of a few days (WSBFa). If these companions are indeed M-dwarfs this would further add to the argument that low mass objects can survive in the outer regions of 47 Tuc for appreciable times. 

Our null result therefore suggests that system metallicity is the dominant factor inhibiting planet formation in 47 Tuc, and is consistent with the expected number of detected planets around stars of cluster metallicity from radial velocity searches (Fig.\ref{expnumbers}). However, 47 Tuc is an old system, and \citet{VM2003} have shown that the transiting planet HD209458b seems to be evaporating its upper atmosphere to a significant degree. Could it be that all the planets have evaporated? These authors conclude that the evaporation timescale is comparable to the main sequence lifetime of the host star, and these planets would then become detectable as hydrogen depleted objects. \citet{Lec2004} give a range of evaporation rates for HD209458b. Even the highest of these (5$\times$10$^{11}$g/s$^{-1}$) gives an evaporation timescale of $\sim$8$\times$10$^{10}$yrs (assuming the rate remains constant), which is $\sim$5.9 times the age of 47 Tuc. 

To display our null result graphically, Fig.\ref{sigmaplot} shows the statistical significance of the result for a variety of different Hot Jupiter occurrence rates (fraction of stars with Hot Jupiter planets), plotted against orbital period. The currently accepted Hot Jupiter occurrence rate for the Solar Neighbourhood \citep{Butler2000,Basri04} is plotted as a dotted line. We can hence rule out Hot Jupiter planets in 47 Tuc to 4$\sigma$ significance to P$=$3.6d, 3$\sigma$ significance to a period of 5.5 days and 2$\sigma$ to P$=$9.2d, if the cluster has the same occurrence rate as the solar neighbourhood. From Fig.\ref{sigmaplot}, any occurrence rate hypothesis from 0.001 to 0.1 can be ruled out in 47 Tuc to a significance of 1$\sigma$$-$$>$5$\sigma$ to a variety of orbital period limits.

\section{Summary and Conclusions}
In this paper, we have presented results of a wide-field ground based search for Hot Jupiter transits in the globular cluster 47 Tucanae. The dataset in total included time series lightcurves (produced via difference imaging) of 21,920 main sequence stars of suitable photometric accuracy to search for planetary companions. The sampled orbital period range was 1-16d, typical of Hot Jupiters found in the solar neighbourhood and defined by the length of our observing window. Seven planets were expected in the dataset, if the Hot Jupiter formation probability is the same as that in the solar neighbourhood, but none were found. This is a null result of high statistical significance (3.3$\sigma$), as tested with extensive Monte Carlo simulations on our dataset and is in agreement with the results of \citet{Gil2000} in an independant and complementary HST study of a region centered on the core of the cluster. 

The main aim of this work was to discriminate between stellar crowding and system metallicity as driving factors for the formation (or otherwise) of Hot Jupiter planets. By observing the whole visible extent of the cluster, (approaching 60$\%$ of the cluster tidal radius) and hence being more sensitive to regions of significantly less crowding than that of \citet{Gil2000}, our results suggest that system metallicity is the dominant factor determining the formation of Hot Jupiter planets in this environment. This result confirms the work of \citet{Gon1997,L2000} and \citet{Santos2001}, who note that extrasolar planets are preferrentially associated with stars of supersolar metallicity, and shows that planetary formation (at least for Hot Jupiters) is not a universal process, being strongly dependent on environment.

\section*{Acknowledgments}
\noindent DW would like to thank the following people for helping make this paper a reality: Brian Schmidt for allowing use of his astrometric package and the help in running it, Ron Gilliland for providing the isochrone data to determine the target star parameters, Antoine Bouchard for the help with IDL and finally the MSSSO TAC for the generous allocation of 40$''$ time. DW would like to thank Ron Gilliland for useful remarks while acting as referee.

\clearpage

\plotone{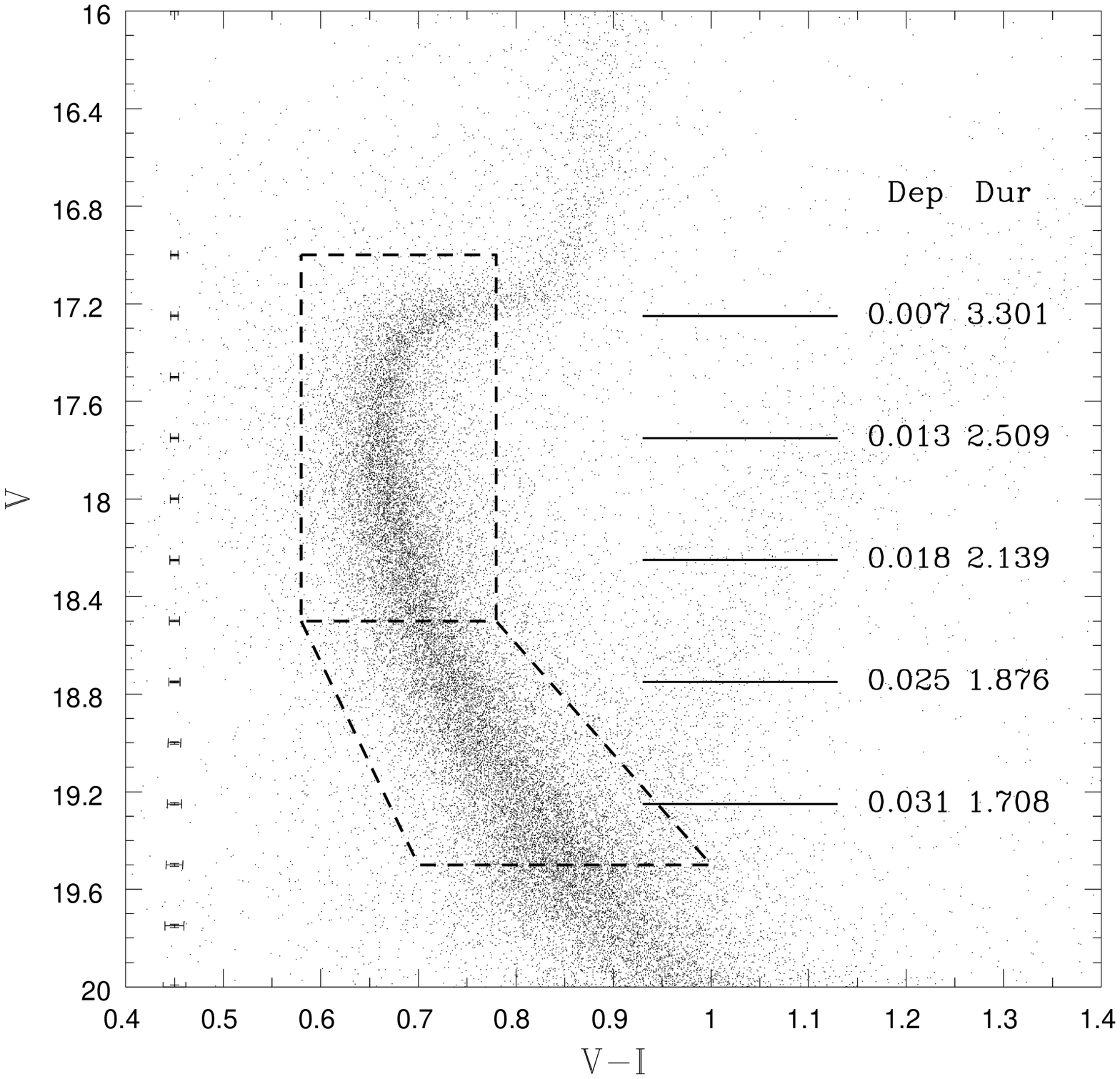}
\figcaption[47Tuc_tr.eps]{The expected depth and duration of planetary transits in 47 Tucanae involving main sequence stars. The predicted depths assume the planet passes on the `average' chord length of the stellar disk, and has a period of 3.3d, with a radius of 1.3R$_J$. The target range for this project is V=17.0-19.5, and hence has corresponding transit depths of 0.007 to 0.031 magnitudes. The two defined boxes show the position of stars  with rms photometric uncertainty $\le$0.02 mag (top), and stars rms $\le$0.04 mag (bottom), and were the two regions searched with the Matched Filter Algorithm. The uncertainly in our CMD dataset is also plotted as error bars.\label{47Tuc_tr_depths}}

\clearpage

\plotone{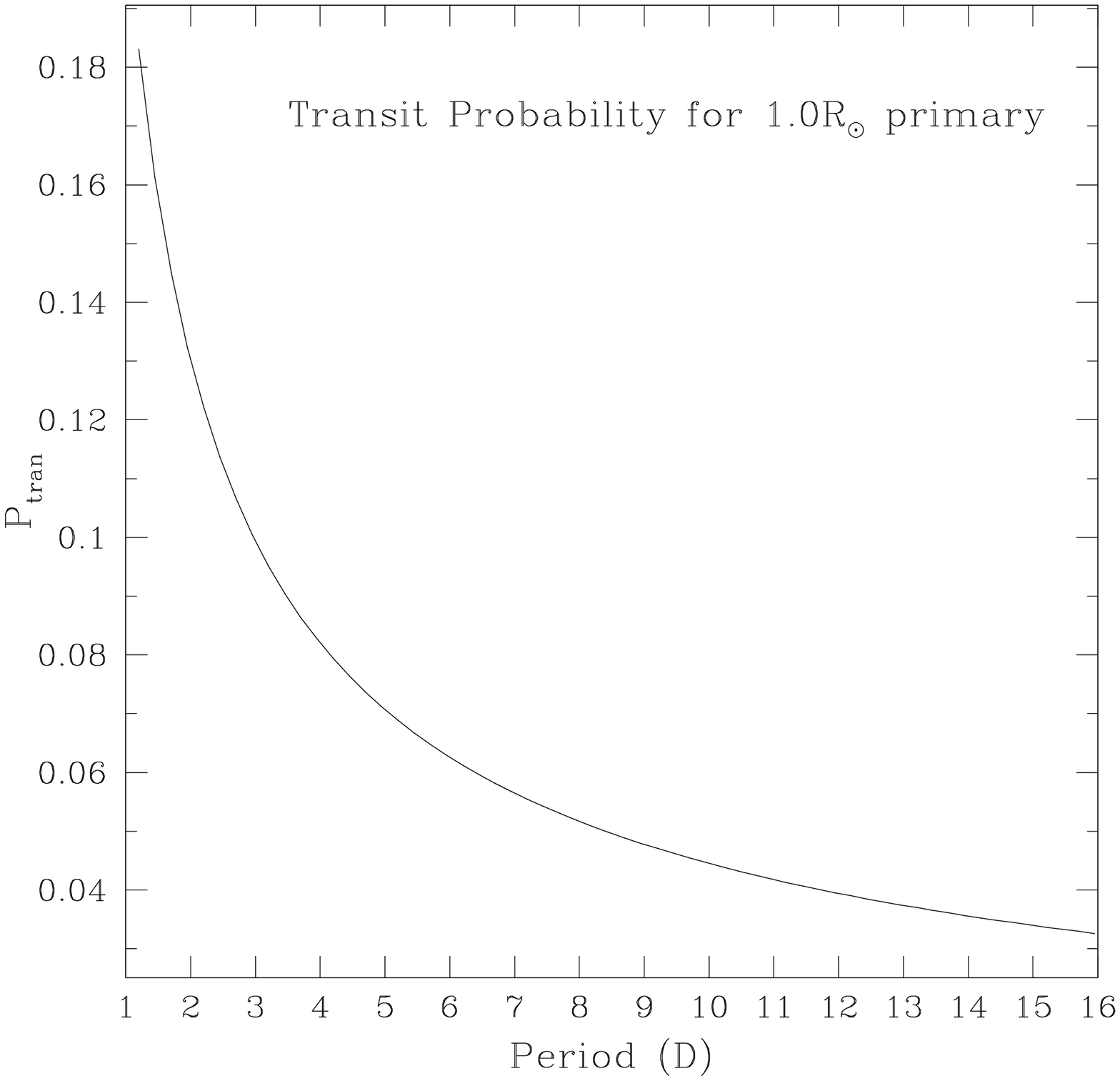}
\figcaption[transprobs.eps]{The geometrical transit probability, P$_{tran}$ for a solar radius primary in 47 Tuc, located in the middle of our transit search range, orbited by a 1.3R$_{J}$ radius secondary. The probability decreases as period increases, and has been plotted to an upper limit of 16d, as a planet of this period could display three detectable (and hence confirmed) transits inside our observing window. By altering the assumed radius to 1R$_{J}$ the resultant transit recoverability levels and expectations were not changed by any significant degree.\label{transprobs}}

\clearpage

\plotone{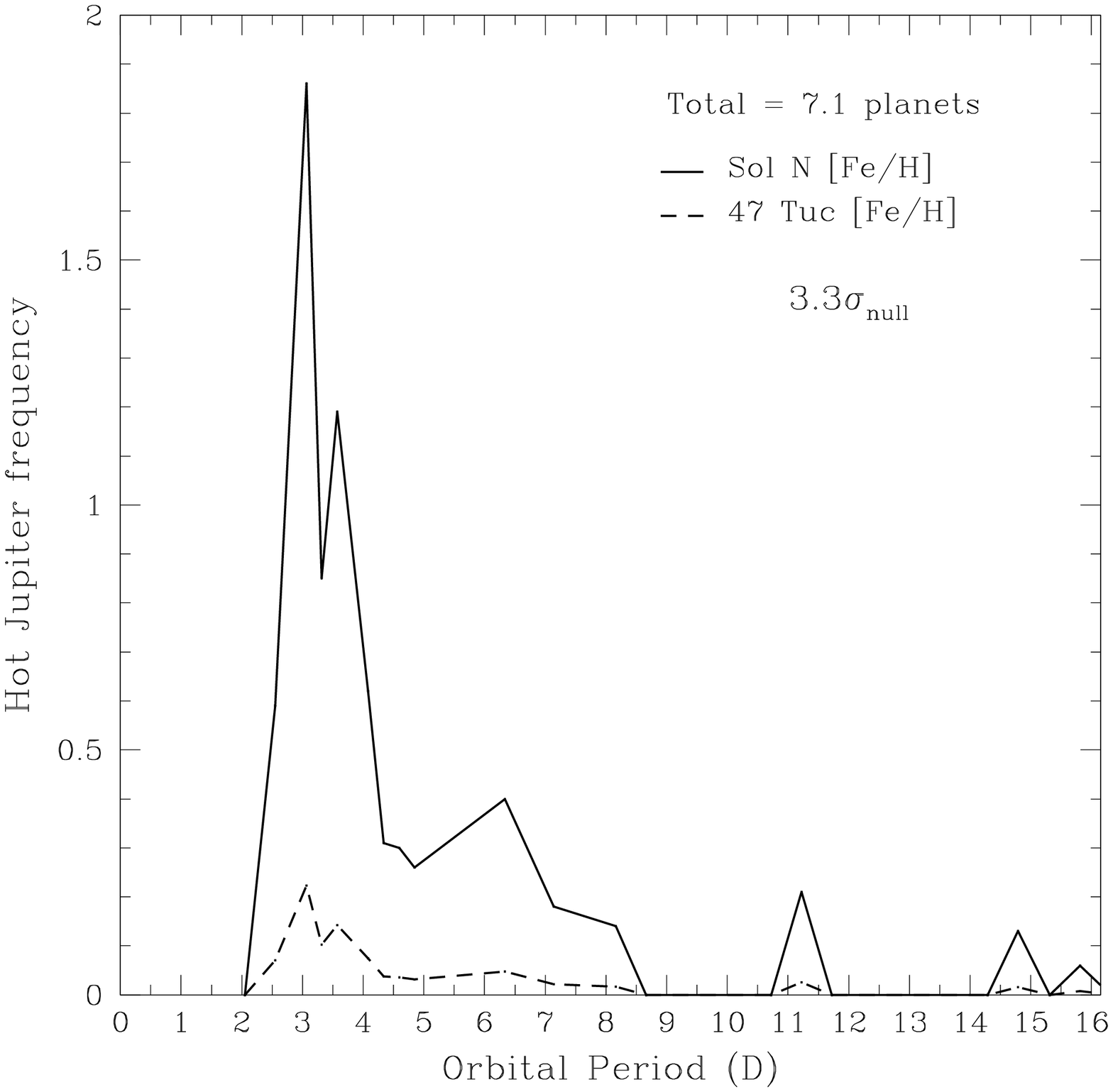}
\figcaption[expnumbers.eps]{The number of expected Hot Jupiter planets detectable in our dataset, as a function of orbital period. These numbers incorporate the Monte Carlo simulation transit recoverability results, and are scaled with the distribution of Hot Jupiter planets in the Solar Neighbourhood. The dashed line is the expected number of planets for stars in the solar neighbourhood at the metallicity of 47 Tuc, totalling 0.9 planets.\label{expnumbers}}

\clearpage

\plotone{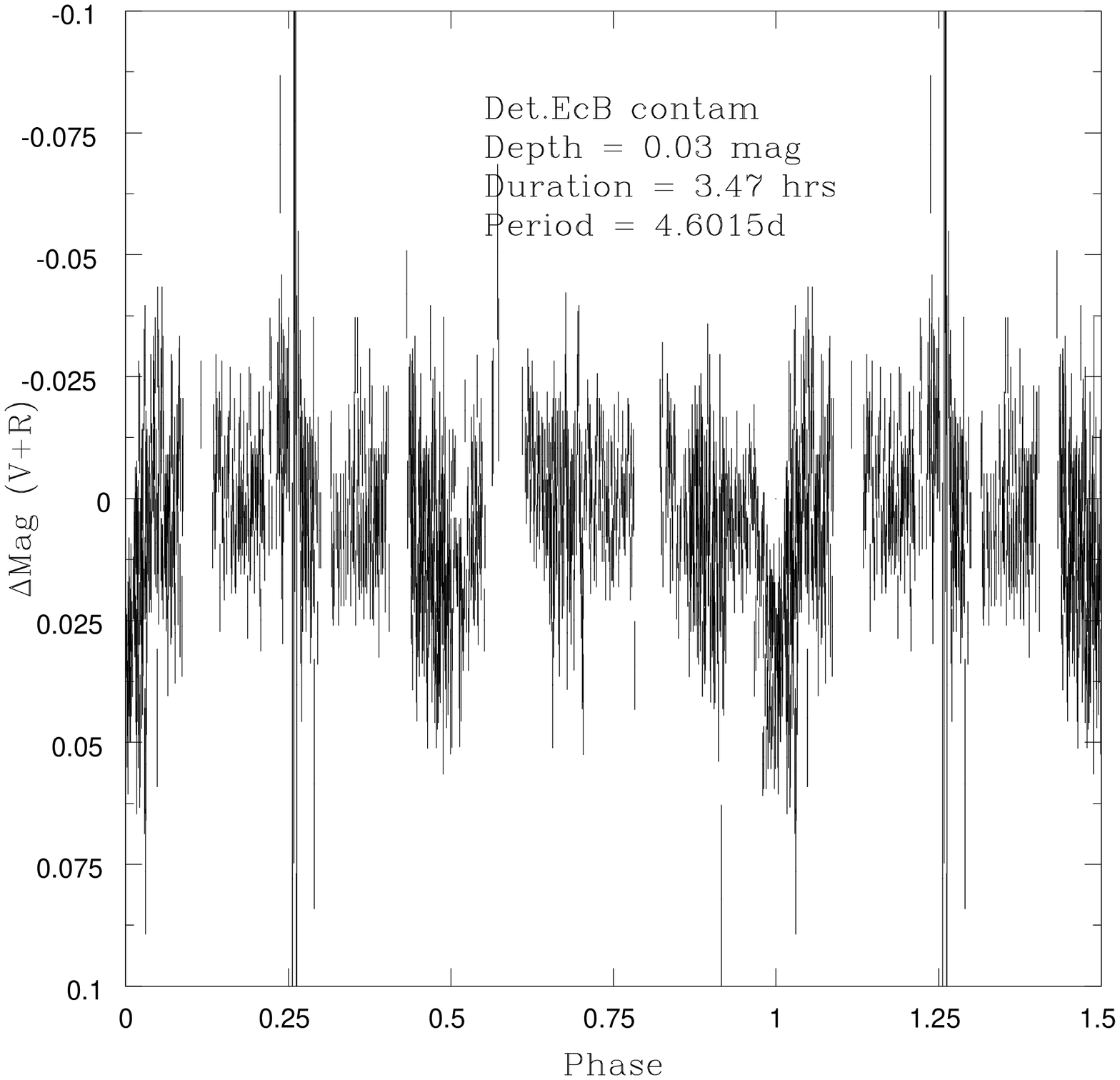}
\figcaption[contamEcB.eps]{The highest significance detection in the transit search for stars $\le$0.02 mag rms. This is contamination of a neighbouring detached eclipsing binary imposed on a 47 Tuc main sequence star, and is a good example of the ability of the MFA to detect transit-like features. The main eclipse occurs at $\Phi$$=$0 and 1. The secondary eclipse can be seen at $\Phi$$=$0.5. The photometric errorbars have been overplotted.\label{contamEcB}}

\clearpage

\plotone{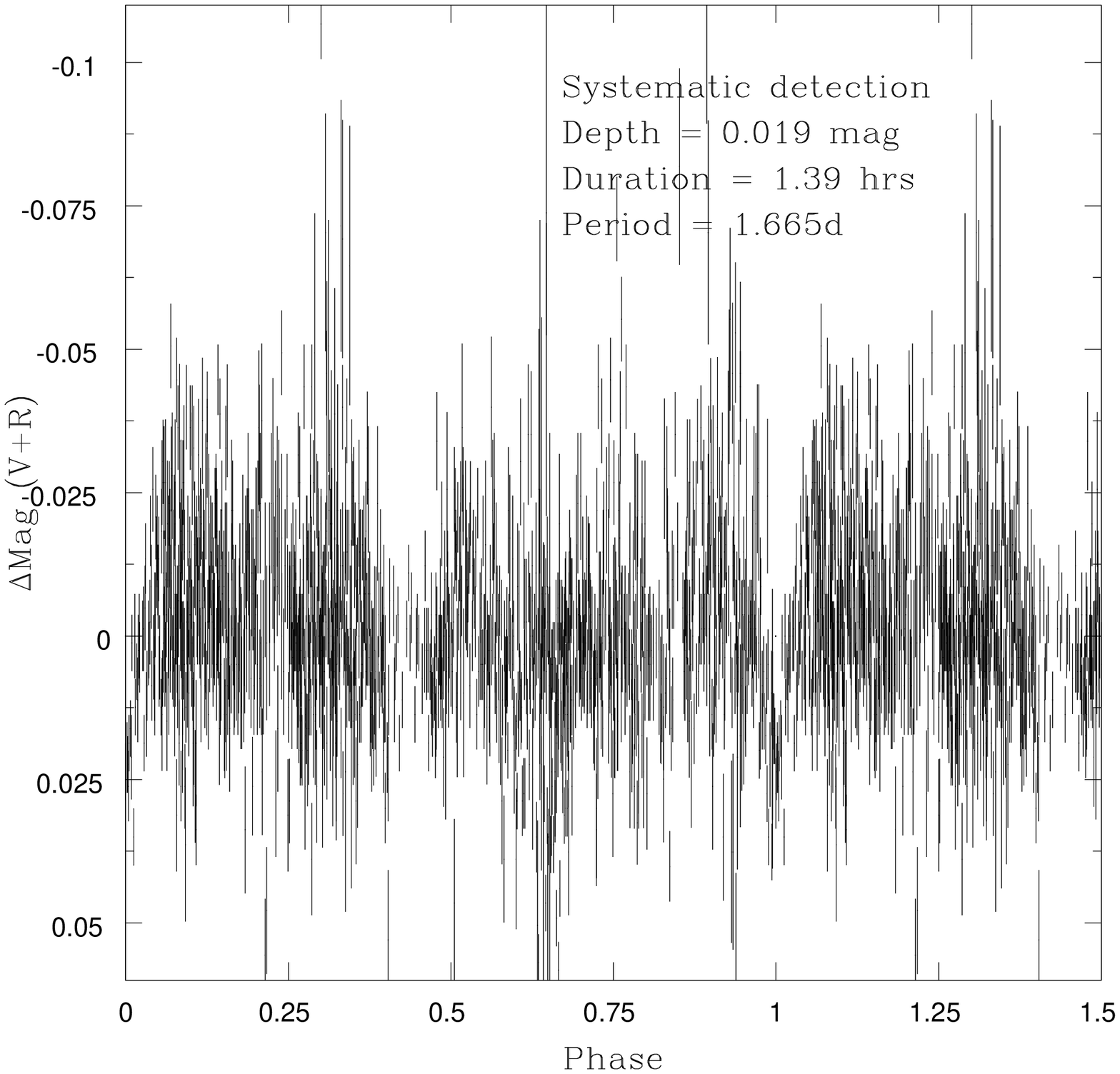}
\figcaption[sysdet.eps]{A typical detection which occurs at a common periodicity and common position in the time-series (at $\Phi$=1) in the $\le$0.02 mag rms sample. This detection has the expected depth of a Hot Jupiter companion, but only $\sim$0.5 the duration. This example shows that a similarly deep transit but with twice the duration would be easily visible in the data. As this derived orbital period was common in the dataset, along with the MJD position of the `transit' features, this has been classified as a systematic effect.\label{sysdet}}

\clearpage

\plotone{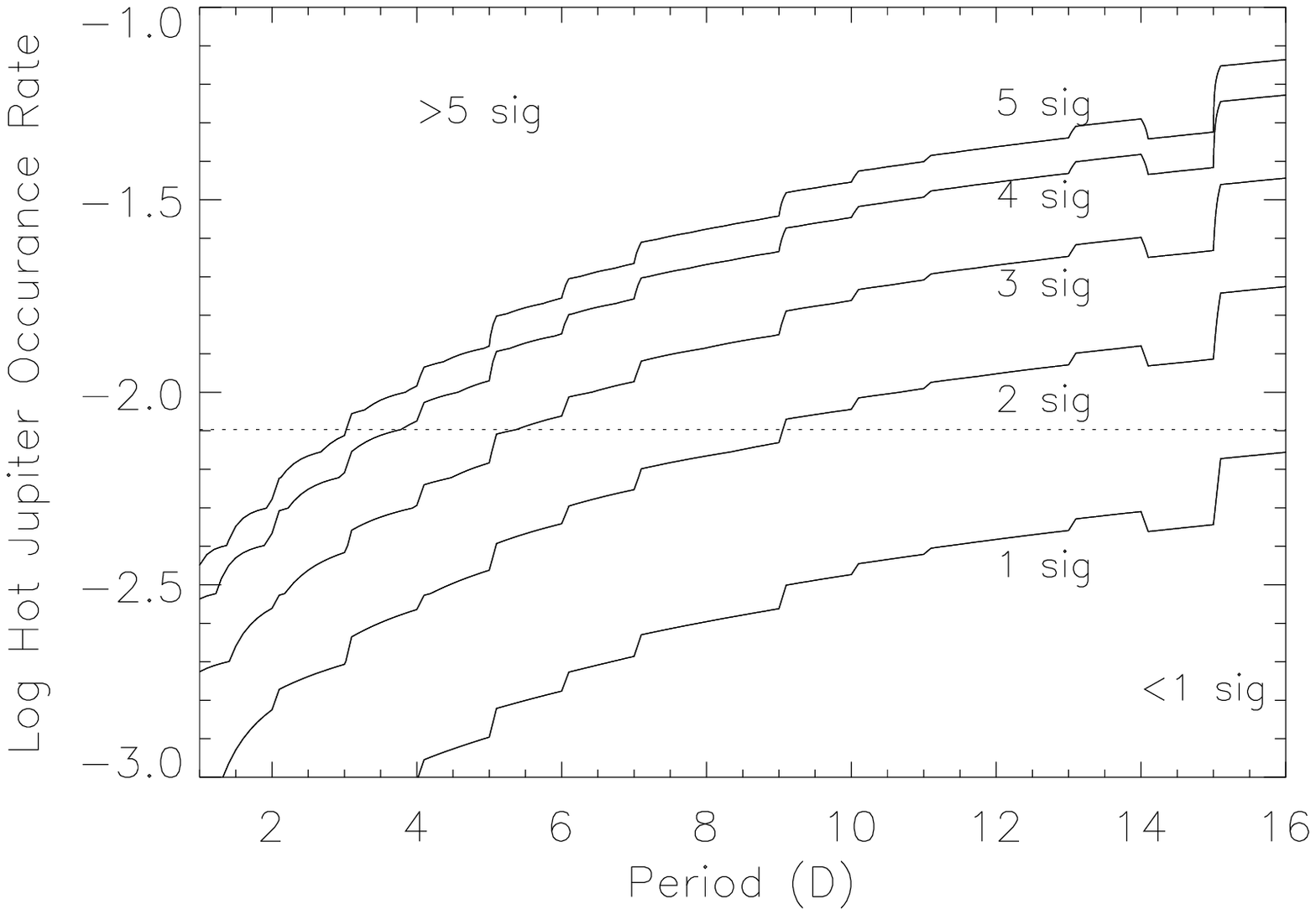}
\figcaption[sigmaplot.eps]{To highlight the null result in a general way, this figure shows the statistical significance that can be placed on a range of Hot Jupiter occurrence rates (fraction of stars with Hot Jupiters) plotted against orbital period for 47 Tuc. The Hot Jupiter occurrence rate in the solar neighbourhood is plotted as a dotted line. Any Hot Jupiter occurrence rate hypothesis can be ruled out to the significance marked to the respective orbital period limit.\label{sigmaplot}}

\clearpage

\begin{center}
\footnotesize
\begin{tabular}{llll} \hline\hline
\noalign{\medskip}
$m_{V}$ & $M_{V}$ & $Mass_{\odot}$ & $Radius_{\odot}$ \\
\noalign{\medskip}
\hline
\noalign{\medskip}
17.0 & 3.60 & 0.923 & 2.347 \\
17.25 & 3.85 & 0.907 & 1.520 \\
17.5 & 4.10 & 0.892 & 1.289 \\
17.75 & 4.35 & 0.875 & 1.142 \\
18.0 & 4.60 & 0.857 & 1.045 \\
18.25 & 4.85 & 0.837 & 0.959 \\
18.5 & 5.10 & 0.814 & 0.885 \\
18.75 & 5.35 & 0.792 & 0.826 \\
19.0 & 5.60 & 0.769 & 0.777 \\
19.25 & 5.85 & 0.746 & 0.737 \\
19.5 & 6.10 & 0.722 & 0.702 \\
\noalign{\medskip}
\hline\hline
\end{tabular}
\end{center}
\vspace*{3mm}
 \normalsize{The parameters of 47 Tuc stars suitable for the transit search, as derived from the isochrones and luminosity functions of \citet{VB2001}, kindly provided by R.Gilliland (private communication). The assumed 47 Tuc distance modulus is 13.4, as derived by those authors.}

\clearpage

\begin{center}
\footnotesize
\begin{tabular}{lllll} \hline\hline
\noalign{\medskip}
Central $Period (D)$ & $F_{HJ}$$_{i}$ & $P_{Tran}$ & $R$ & $N_{HJ}(P)$ \\
\noalign{\medskip}
\hline
\noalign{\medskip}
2.552 & 7.29 & 0.11 & 0.73 & 0.59  \\
3.062 & 29.17 & 0.098 & 0.65 & 1.86 \\
3.317 & 14.58 & 0.092 & 0.63 & 0.85 \\
3.572 & 21.88 & 0.088 & 0.62 & 1.19 \\
4.082 & 14.58 & 0.080 & 0.53 & 0.62 \\
4.337 & 7.29 & 0.077 & 0.56 & 0.31 \\
4.592 & 7.29 & 0.076 & 0.54 & 0.30 \\
4.847 & 7.29 & 0.072 & 0.49 & 0.26 \\
6.377 & 14.58 & 0.060 & 0.46 & 0.40 \\
7.142 & 7.29 & 0.056 & 0.44 & 0.18 \\
8.162 & 7.29 & 0.050 & 0.38 & 0.14 \\
11.222 & 14.58 & 0.042 & 0.34 & 0.21 \\
14.792 & 14.58 & 0.034 & 0.27 & 0.13 \\
15.812 & 7.29 & 0.033 & 0.25 & 0.06 \\
\noalign{\medskip}
\hline\hline
\end{tabular}
\end{center}
\vspace*{3mm}
 \normalsize{The expected number of detectable Hot Jupiter planets in the dataset as a function of orbital period P, $N_{HJ}(P)$, as described in section 3.2. The total number of expected detctions is 7 planets, assuming the formation probability of Hot Jupiters is the same in 47 Tuc as it is in the solar neighbourhood.}

\end{document}